\documentstyle{article}
\begin{document}
\newcommand{\sss}[1]{\mbox{\scriptsize #1}}
\newcommand{\SL}{{\cal S}_{\sss{L}}}
\newcommand{\SNL}{{\cal S}_{\sss{NL}}}
\newcommand{\Tr}{\mbox{Tr}\,}
\newcommand{\TF}{\mbox{\boldmath $F$}}
\renewcommand{\arraystretch}{1.1}
\newcommand{\flin}[2]{\ArrowLine(#1)(#2)}
\newcommand{\glin}[3]{\Photon(#1)(#2){2}{#3}}
\newcommand{\nl}{\nonumber \\}
\newcommand{\lsim}
{\mathrel{\raisebox{-.3em}{$\stackrel{\displaystyle <}{\sim}$}}}
\newcommand{\sof}{\SetOffset}
\newcommand{\vj}[4]{{\sl #1~}{\bf #2 }
 \ifnum#3<100 (19#3) \else (#3) \fi #4}
\newcommand{\ej}[3]{{\bf #1~}
 \ifnum#2<100 (19#2) \else (#2) \fi #3}
\newcommand{\ap}[3]{\vj{Ann.~Phys.}{#1}{#2}{#3}}
\newcommand{\app}[3]{\vj{Acta~Phys.~Pol.}{#1}{#2}{#3}}
\newcommand{\cmp}[3]{\vj{Commun. Math. Phys.}{#1}{#2}{#3}}
\newcommand{\cnpp}[3]{\vj{Comments Nucl. Part. Phys.}{#1}{#2}{#3}}
\newcommand{\cpc}[3]{\vj{Comp. Phys. Commun.}{#1}{#2}{#3}}
\newcommand{\epj}[3]{\vj{Eur. Phys. J.}{#1}{#2}{#3}}
\newcommand{\hpa}[3]{\vj{Helv. Phys.~Acta}{#1}{#2}{#3}} 
\newcommand{\ijmp}[3]{\vj{Int. J. Mod. Phys.}{#1}{#2}{#3}}
\newcommand{\jetp}[3]{\vj{JETP}{#1}{#2}{#3}}
\newcommand{\jetpl}[3]{\vj{JETP Lett.}{#1}{#2}{#3}}
\newcommand{\jmp}[3]{\vj{J. Math. Phys.}{#1}{#2}{#3}}
\newcommand{\jp}[3]{\vj{J. Phys.}{#1}{#2}{#3}}
\newcommand{\lnc}[3]{\vj{Lett. Nuovo Cimento}{#1}{#2}{#3}}
\newcommand{\mpl}[3]{\vj{Mod. Phys. Lett.}{#1}{#2}{#3}}
\newcommand{\nc}[3]{\vj{Nuovo Cimento}{#1}{#2}{#3}}
\newcommand{\nim}[3]{\vj{Nucl. Instr. Meth.}{#1}{#2}{#3}}
\newcommand{\np}[3]{\vj{Nucl. Phys.}{#1}{#2}{#3}}
\newcommand{\pl}[3]{\vj{Phys. Lett.}{#1}{#2}{#3}}
\newcommand{\prp}[3]{\vj{Phys. Rep.}{#1}{#2}{#3}}
\newcommand{\pr}[3]{\vj{Phys. Rev.}{#1}{#2}{#3}}
\newcommand{\prl}[3]{\vj{Phys. Rev. Lett.}{#1}{#2}{#3}}
\newcommand{\ptp}[3]{\vj{Prog. Theor. Phys.}{#1}{#2}{#3}}
\newcommand{\rpp}[3]{\vj{Rep. Prog. Phys.}{#1}{#2}{#3}}
\newcommand{\rmp}[3]{\vj{Rev. Mod. Phys.}{#1}{#2}{#3}}
\newcommand{\rnc}[3]{\vj{Rivista del Nuovo Cim.}{#1}{#2}{#3}}
\newcommand{\sjnp}[3]{\vj{Sov. J. Nucl. Phys.}{#1}{#2}{#3}}
\newcommand{\sptp}[3]{\vj{Suppl. Prog. Theor. Phys.}{#1}{#2}{#3}}
\newcommand{\zp}[3]{\vj{Z. Phys.}{#1}{#2}{#3}}
\newcommand{\jop}[3]{\vj{Journal of Physics} {\bf #1} (#2) #3}
\newcommand{\ibid}[3]{\vj{ibid.} {\bf #1} (#2) #3}
\newcommand{\hep}[1]{{\sl hep--ph/}{#1}}
\begin{center}
{\Large \bf Theoretical issues at LEP2 and LC} \\
\vspace{4mm}
Roberto Pittau\\
Dipartimento di Fisica Teorica, 
Universit\`a di Torino, Italy\\
INFN, Sezione di Torino, Italy
\end{center}

\begin{abstract}
After 12 years of glorious data taking LEP has been shut down. 
During the past two years an incredible effort has been
devoted to get more accurate predictions and estimates of the
related theoretical uncertainties \cite{wshop}. 
Many of the theoretical questions
driven by LEP are also relevant at the LC, where predictions will
be needed with even higher accuracy.
This is particularly true for 4-fermion Physics. 
In this contribution, I review some of the most important 
theoretical achievements at LEP in understanding $W$ and $Z$ 
pair production, 4-fermion + 1 visible photon signatures
and in solving problems related with gauge invariance.
Issues on single $W$ Physics are covered in reference \cite{acco}.
 Part of the presented work is in progress, the 
final LEP2 analysis still being under way.
\end{abstract}

\section*{$W$-pair production} When collecting 
data at $\sqrt{s}=$ 189 GeV, a deficit 
in the number of events was observed at LEP2, with respect 
to the Standard Model predictions. 
This fact triggered a re-analysis of the
available tools for calculating the total cross section $\sigma_{WW}$.
At that time, a theoretical 2\% error band was assigned to this observable, 
two times bigger than the experimental error.
Therefore, it was immediately clear that the computation of the genuine 
Electro Weak effects was needed to match the experimental accuracy. 
On the other hand a full four-fermion one-loop EW calculation 
was (and still is) beyond reach, and including only the $WW$-like 
diagrams violates gauge invariance.
 The solution to this problem is represented by the so called Double 
Pole Approximation (DPA) \cite{dpa}.
The DPA isolates the poles at the complex squared 
masses, with gauge invariant residues which are then projected 
onto the respective on-shell gauge invariant counterparts.
The projection is from the off-shell phase space to the 
on-shell phase space. Even though such a procedure is strictly gauge invariant,
the projection procedure is not unique. However, 
the ambiguity is small, namely 
${\cal O}(\frac{\alpha}{\pi}\frac{\Gamma_W}{M_W})$.

 Applying DPA to $W$-pair production means that only the double-pole 
residues of the two resonances are considered, 
and one-loop EW contributions included there,
for which only (available) on-shell corrections are needed.
The corrections to be included fall in two different classes, namely
factorizable contributions, in which the production, propagation and decay
steps are clearly separated, and non-factorizable contributions, in which
a photon with energy $E_\gamma \lsim \Gamma_{W}$ is emitted.

The DPA is not reliable at the $W$-pair threshold, where  
the background diagrams get important.
The expected DPA uncertainty above threshold
is of the order

\noindent  ${\cal O}\left(\frac{\alpha}{\pi}\frac{\Gamma_W}{M_W} 
{\rm ln( \cdots)}\right)\,< 0.5\%$, in fact, when
$\sqrt{s} > 2\,M_W + n \Gamma_W$ with $n= {\cal O}(3-5)$,  
the background diagrams are of the order
$\sim \frac{\alpha}{\pi}\frac{\Gamma_W}{\sqrt{s}-M_W}\, 
{\rm ln( \cdots)} \sim 0.1\%$.

Very far away from resonance, the DPA cannot be used any more.

At LEP2 energies, the inclusion of the DPA formalism in {\tt RACOONWW} 
\cite{raw},  
{\tt BBC} \cite{bbc} and {\tt YFSWW} \cite{kow} 
allows to lower the theoretical uncertainty 
on $\sigma_{WW}$ from 2\% to 0.5 \%, in much better agreement with the data.

In conclusion, with the help of the DPA, a theoretical 
accuracy at the level of 0.5 \% on $\sigma_{WW}$ is reached, as 
required by the LEP2 collaborations \cite{wshop}.
The error decreases with increasing energy, giving the following 
estimates of the theoretical uncertainty
on $\sigma_{\rm WW}$
\begin{itemize}
\item[ ] 0.4 \% at $\sqrt{s}= 200$ GeV,~
         0.5 \% at $\sqrt{s}= 180$ GeV,~
         0.7 \% at $\sqrt{s}= 170$ GeV.
\end{itemize}
A theoretical uncertainty of the order of 1 \% must be assigned to 
the distributions.

One of the basic differences between {\tt YFSWW} and {\tt RACOONWW} is the
treatment of the photon radiation. The former code uses {\tt YFS}
techniques to include multi-$\gamma$ emission, while the latter 
contains the full matrix element for the real emission of 1 hard photon.
 As a matter of principle the Leading Log approximation used in
{\tt YFSWW} is only valid is kinematical regions where the Logarithms
are really leading, while {\tt RACOONWW} misses multiple photon emission.
Thus the problem of correctly estimating the size of the neglected 
effects remains open, at least until when a full
one-loop calculation will become available.
This will be required for high precision measurements at the LC.

\section*{Preserving gauge invariance}
The problem is  re-summing propagators of instable particles 
in a gauge invariant way.
Several solutions are available. In the Complex Mass Scheme (CMS) \cite{raw}
complex bosonic masses are used everywhere, also in the definition of
$\sin^2 \theta_W$. While the Fermion Loop approach (FL) \cite{fl}
takes 
into account the Imaginary part of all fermion-loop diagrams.
In the Exact Fermion Loop (EFL) \cite{efl}, also the real contributions are
included, that are a numerically important ingredient of the full 
one-loop answer.

 In spite of its simplicity, CMS is unrealistic in the sense that
instable particles acquire a width irrespective of the fact that
the flowing 4-momentum squared is space-like or time-like.
On the other hand the FL solutions cannot be used for particles with
non-fermionic decay modes.
 A third solution, the so called Non Local Approach (NLA) \cite{nla}, 
has been proposed, that consists of adding to the 
SM action gauge invariant non local pieces, by using a path-ordered 
exponential $U_2$ that carries the gauge transformation from one 
space-time point to the other:
\begin{eqnarray}
  \SNL^{\sss{YM}} &=& -\,\frac{1}{4} \int d^4x\,d^4y\,\Sigma_1(x-y)\,
           B_{\mu\nu}(x)\,B^{\mu\nu}(y) \nonumber \\[1mm]
                  & & -\,\frac{1}{2} \int d^4x\,d^4y\,\Sigma_2(x-y)\,\Tr\Bigl[ 
           U_2(y,x)\,\TF_{\mu\nu}(x)\,U_2(x,y)\,\TF^{\mu\nu}(y) \Bigr]
       \nonumber \\
  \SNL^{\Phi}     &=& -\,\frac{g_1g_2}{2M_W^2} \int d^4x\,d^4y\,\Sigma_3(x-y)\,
                  [ \Phi^{\dagger}(x)\,\TF_{\mu\nu}(x)\,\Phi(x) ]\,
                  B^{\mu\nu}(y) \nonumber \\
              & & -\,\frac{g_2^4}{4M_W^4} \int d^4x\,d^4y\,\Sigma_4(x-y)\,
                  [ \Phi^{\dagger}(x)\,\TF_{\mu\nu}(x)\,\Phi(x) ]\,
                  [ \Phi^{\dagger}(y)\,\TF^{\mu\nu}(y)\,\Phi(y) ]\,.
\label{eq1}
\end{eqnarray}
The resulting Lagrangian can then be used to derive the needed
Feynman Rules for the transverse self-energies plus extra vertices, 
therefore preserving 
gauge invariance by construction.
$\Sigma_{1,..,4}$ can be arbitrary but one can also insert   
the computed SM self-energies in equation (\ref{eq1}) to match 
the effective Lagrangian to the 
Standard Model. Both approaches are implemented in 
{\tt NEXTCALIBUR} \cite{ner} 
and 
a few results are presented in tables \ref{tab1} and \ref{tab2}.
%--
\begin{table}[thb]
\begin{center}
\begin{tabular}{|c|c|c|}
\hline
$\sqrt{s}=200$ GeV & $\sigma(\mu \nu_\mu u \bar d) $ [pb] 
 & $\sigma(e \nu_e u \bar d)$  [pb]\\
\hline
{\tt NEXTCALIBUR\_CMS}  & 0.6945(55)& 0.8791(83)  \\
\hline
{\tt NEXTCALIBUR\_NLA} & 0.6946(55)& 0.8792(83)  \\
\hline
\end{tabular}
\end{center}
\caption{\label{tab1} CMS vs NLA schemes as implemented 
in {\tt NEXTCALIBUR}. $\Sigma_{1,..,4}$ are such that the
{\em running width} parameterization for the propagators
$P^{-1}(s) = (s-M^2+i\,s \Gamma/M)~(s>0)$ is reproduced.}
\end{table}
%--
\begin{table}[thb]
\begin{center}
\begin{tabular}{|c|c|}
\hline
$\sqrt{s}=200$ GeV & $\sigma(\mu \nu_\mu u \bar d) $ [pb] \\
\hline
{\tt NEXTCALIBUR\_CMS} & 0.6849(39)\\
\hline
{\tt NEXTCALIBUR\_NLA} & 0.6847(39)\\
\hline
\end{tabular}
\end{center}
\caption{\label{tab2} CMS vs NLA schemes as implemented 
in {\tt NEXTCALIBUR}. $\Sigma_{1,..,4}$ are computed
in the SM and their imaginary part at the peak is used 
to produce the CMS numbers.}
\end{table}

When one is interested in the low energy regime 
(e.g. Single-W processes), non local longitudinal Operators have 
to be considered as well. Then, one can match the results
of the EFL approach \cite{pre}.
\section*{$Z$-pair production}
Less accuracy is required at LEP2 for this observable with respect 
to the $W$-pair case.
 Thought feasible in principle, a DPA $ZZ$ calculation 
is not available yet. A theoretical accuracy of 2\% on
$\sigma_{ZZ}$ is at present estimated at LEP2 by varying
the renormalization scheme and by comparing ${\tt YFSZZ}$ \cite{yfz} and 
${\tt ZZTO}$ \cite{zzo}.
{\tt YFS} photon exponentiation is used by the former code, while the latter
includes Exact Fermion Loop corrections.

\section*{Four fermions plus 1 visible photon} 
This signature gives information on the quartic gauge coupling
and is relevant when studying processes with three final state bosons, such as
$W^+ W^- \gamma$ production, $Z Z \gamma$ and $Z \gamma \gamma$.
Furthermore, it is a building block for the full computation of
$e^+ e^- \to$ 4f at ${\cal O}({\alpha})$.
A bunch of codes contributed, with different strategies.
{\tt CompHEP} \cite{cop}, {\tt GRACE} \cite{gre} and 
{\tt HELAC} \cite{phs} compute the 
exact Matrix Element (ME) with massive fermions.
{\tt RACOONWW} uses exact ME, but in the limit of massless fermions.
{\tt NEXTCALIBUR} generates photons only through $p_t$ dependent SF.
{\tt WRAP} \cite{swp} has a matching between ME and SF generated photons. 
The last approach allows to estimate the size of the double 
counting when blindly dressing the $4f + \gamma$ ME 
with collinear ISR.  {\tt WRAP} observed effects up to 5\%, depending
on the energy cut used to define the visible photon.

In conclusion, a very good technical precision has been reached
in the computation of four-fermion processes plus 1
additional photon. However, the 
non-logarithmic ${\cal O}(\alpha)$ corrections
are not known. Therefore a 2.5\% theoretical accuracy 
on total cross section and inclusive distributions is estimated
at LEP2 energies. Larger effects are expected at the LC.

\end{document}